\documentclass[11pt]{article} 
\usepackage{hyperref} 
\pdfoutput=1 
	
\begin{document} 
\title{When water freezes before it solidifies} 
\author{Faryar Tavakoli, Alen Sarkissians, Pirouz Kavehpour \\ 
\\\vspace{6pt} Mechanical and Aerospace Engineering Department \\ University of California, Los Angeles
, CA 90024, USA} 
\maketitle 
%% The abstract (in this file, and that submitted as text to arXiv) should 
%% "fluid dynamics video" or "fluid dynamics videos" 
\begin{abstract}
\end{abstract} 

We present a fluid dynamics video which illustrates a kinetics dominated phase change just prior to solidification inception for a static drop situated on a cold solid substrate. This metastable stage initiated from solid-vapor-liquid interface and propagate through drop volume at much faster speeds than growth stage.

\section{Introduction and Observation}
\indent

Most of the liquid solidification research, particularly for water, phase change is assumed to occur at equilibrium freezing temperature; however, this is not always the case especially for liquids that can undergo supercooling. Supercooling is the process of chilling a liquid below its freezing point, without the phase change from liquid to solid. When supercooling of a liquid drop exceeds few C, the process of freezing splits in 4 distinct stages: i) liquid droplet cooling, ii) stage-one freezing (also known as recalescence among metallurgists), iii) stage-two freezing, and iv) solid droplet cooling. The recalescence stage has not been fully understood and its structural morphologies have not been fully visualized for static drops on a cold solid substrate.
\indent
  
Experiments were organized to look closely into location of this kinetics driven phase change (recalescence). For this, different sizes of water drops were situated on either hydrophobic or hydrophilic substrates and cooled by a peltier element to temperatures ranging from 0 to -25$^\circ$C. It should be mentioned that water drops and substrate are cooled simultaneously from 25$^\circ$C to designated temperatures. No solidification occurred when water drops were cooled from 0 to -10$^\circ$C because of water ability to undergo supercooling. The movie shows the recalescence stage from top and side views at multiple recording speeds. In all the experiments, nucleation occurs preferentially at trijunction point, supporting energetic arguments that a solid nuclei at the contact line leads to a reduction in the nucleation barrier energy. No immersion mode nucleation (within drop volume) was observed during the experiments.8 After nucleation, recalescence front propagates through drop volume radially in different formations. In the early stages of recalescence, smooth surface is formed and remains planar until the interface is subject to unstable growth which leads to various patterns such as cellular, dendritic and fractal morphologies. 

\end{document}